\documentstyle[12pt]{article}
\renewcommand{\d}{\delta}

\newcommand{\QFT}{{\rm QFT}}

\newcommand{\D}{\Delta}

\newcommand{\FT}{{\rm FT}}

\newcommand{\ar}{\longrightarrow}

\newcommand{\s}{\sigma}
\newcommand{\la}{\lambda}

\begin{document}
\title{Implementation of Quantum Fourier Transform and Simulation of Wave
 Functions by Fixed Interaction}
\author{Yuri Ozhigov
\thanks{Institute of Physics and Technology, Moscow, e-mail address: 
ozhigov@ftian.oivta.ru.}}
\date{}
\maketitle
\begin{abstract} We study a quantum computer with fixed and permanent 
interaction of diagonal type between qubits. It is controlled only by one-qubit 
quick transformations. It is shown how to implement Quantum Fourier Transform 
and to solve Shroedinger equations with linear and quadratic potentials by a
 quantum computer of such type. The method is adaptable to the wide range of 
interactions of diagonal form between qubits and to the case when different 
pairs of qubits interact variously.
\end{abstract}
\section{Introduction}

The main difficulty in practical implementation of quantum computing is to 
fulfill two qubits transformations playing a crucial role in quantum algorithms. 
To perform such transformations specially we must in fact artificially and 
exactly control the degree of their entanglement that is determined by 
overlapping of their spatial wave functions. However to distinguish different 
qubits the share of overlapping amplitudes must be much less than the overall 
amount of amplitude and thus in the same degree one-qubits transformations are 
easier to fulfill than two qubits transformations. While we perform a 
transformation with one pair of qubits a physical interaction between other 
pairs of qubits cannot be stopped. This permanent interaction existing in all 
real systems requires special and nontrivial methods of correcting. These 
difficulties complicate a straightforward implementation of quantum algorithms. 
Here we shall study a nonstandard model of quantum computer possessing formally 
more narrow possibilities but which may be more feasible. It is controlled by 
only one-qubit transformations whereas two qubits interactions are fixed and 
determined by the spatial disposition of qubits, they go permanently in course 
of computation. To show the possibilities of such model we shall at first study 
how to implement Quantum Fourier Transform (QFT) by such quantum computer 
assuming that the potential of two qubits interaction has a diagonal form and 
decreases as Yukawa potential. Then it will be shown how this method can be 
generalized to wide range of interactions of diagonal form. By means of properly 
chosen one-qubit transformations this method can be easily adopted to the case 
when different pairs of qubits interact differently. At last this approach will 
be applied to the solution of Shoedinger equation for linear and quadratic 
potentials. 

Quantum Fourier Transform (QFT) is a key subroutine in quantum computing. It is 
used in variety of algorithms (look at \cite{Sh, AL, Oz}) as a main step 
generating interference of amplitudes which makes quantum computations so 
powerful. A simple quantum gate array implementing the reversal for QFT is shown 
at the picture 1. It was proposed in several works and was used by Shor for fast 
factoring (see \cite{Sh}). 
Let us agree to represent an integer of the form $a=a_0 +a_0 2+\ldots +a_{l-1} 
2^{l-1}$ by the basic state $|a_0 \ a_1 \ \ldots\ a_{l-1} \ \rangle =|a\rangle$
 forming a basis for input states of a gate array and dispose all $a_j$ from top 
to bottom. The same agreement will be for output only binary figures $b_j$ for 
an integer $b=b_0 +b_0 2+\ldots +b_{l-1} 2^{l-1}$ will be written in reverse 
order - from bottom to top.

\begin{picture}(500,315)(0,-35)
\multiput(80,50)(0,50){5}{\line (30,0){270}}
\put(95,50){\circle{10}}
\put(125,100){\circle{10}}
\put(170,150){\circle{10}}
\put(230,200){\circle{10}}
\put(310,250){\circle{10}}
\put(110,50){\line(0,1){50}}
\put(140,50){\line(0,1){100}}
\put(155,100){\line(0,1){50}}
\put(185,50){\line(0,1){150}}
\put(200,100){\line(0,1){100}}
\put(215,150){\line(0,1){50}}
\put(245,50){\line(0,1){200}}
\put(260,100){\line(0,1){150}}
\put(275,150){\line(0,1){100}}
\put(290,200){\line(0,1){50}}
\put(110,50){\circle*{3}}
\put(140,50){\circle*{3}}
\put(185,50){\circle*{3}}
\put(245,50){\circle*{3}}
\put(155,100){\circle*{3}}
\put(200,100){\circle*{3}}
\put(260,100){\circle*{3}}
\put(215,150){\circle*{3}}
\put(275,150){\circle*{3}}
\put(290,200){\circle*{3}}
\put(110,100){\circle*{3}}
\put(140,150){\circle*{3}}
\put(155,150){\circle*{3}}
\put(185,200){\circle*{3}}
\put(200,200){\circle*{3}}
\put(215,200){\circle*{3}}
\put(245,250){\circle*{3}}
\put(260,250){\circle*{3}}
\put(275,250){\circle*{3}}
\put(290,250){\circle*{3}}
\put(60,50){$a_4$}
\put(60,100){$a_3$}
\put(60,150){$a_2$}
\put(60,200){$a_1$}
\put(60,250){$a_0$}
\put(360,50){$b_0$}
\put(360,100){$b_1$}
\put(360,150){$b_2$}
\put(360,200){$b_3$}
\put(360,250){$b_4$}
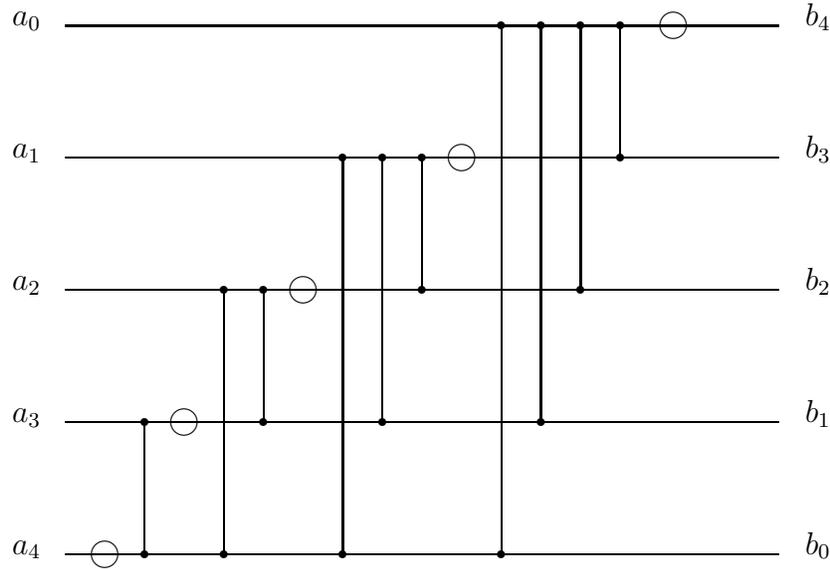
\put(60,10){Picture 1. Gate array for $\QFT^{-1}$ with one and two qubits 
control. Circles}
\put(110,-5){denote Hadamard gates, two qubits gates has the form (\ref{Ham})}
\end{picture}

This array fulfills $\QFT^{-1}$ in $O(l^2 )$ steps whereas its matrix is $N=l^2$ 
dimensioned. 

However, a direct implementation of this scheme requires a control over two 
qubits transformations and thus it does not fit into our model of quantum 
computer. In this paper it is shown how QFT and its reversal can be implemented 
by means of fixed Hamiltonian of two particles interaction of the form 
\begin{equation}
{\rm A)}\ H=\left(
\begin{array}{ccccc}
&0 &0 &0 &0\\
&0 &0 &0 &0\\
&0 &0 &0 &0\\
&0 &0 &0 &\rho
\end{array}
\right)
,\ \ \rho >0, \ \ \ \ \ \ \ \ \ \ \ 
{\rm B)}\ 
H=\left(
\begin{array}{ccccc}
&\rho_1 &0 &0 &0\\
&0 &\rho_2 &0 &0\\
&0 &0 &\rho_3 &0\\
&0 &0 &0 &\rho_4
\end{array}
\right) ,
\label{Ham}
\end{equation}
where all $\rho =\rho_0 \frac{e^{-br}}{r}$; $b=const$; $r$ is a distance between 
the particles and $\rho_1 +\rho_4 \neq \rho_2 +\rho_3$. Dispose our $l$ qubits 
on one line with equal intervals. We shall consider a case when interaction 
between $j$th and $k$th qubits will have Hamiltonian $H_{j,k}$ of the forms 
(\ref{Ham}).

This type of Hamiltonians appears for example in Ising model for particles with 
spin 1/2. The required decrease of interaction with the distance could be 
obtained by placing each particle in the appropriate potential hole. Choosing 
appropriate unit of the length we can make $b=1$. At first we shall study fixed 
interaction of the form (\ref{Ham}, A) and then extend our results to 
(\ref{Ham}, B). 

\section{Implementation of QFT in within phase shifts}

We assume that QFT and its reversal have the form \footnote{This agreement 
corresponds to the definition of ordinary Fourier transform. Typically in 
quantum computing literature it is assumed a reversal definition.}:
\begin{equation}
\QFT :\ |a\rangle\ar\frac{1}{\sqrt{N}}\sum\limits_{b=0}^{N-1}e^{-\frac{2\pi i\ 
ab}{N}}|b\rangle ,
\ \ \ \ \ \ \ \ 
\QFT ^{-1} :\ |a\rangle\ar\frac{1}{\sqrt{N}}\sum\limits_{b=0}^{N-1}e^{\frac{2\pi 
i\ ab}{N}}|b\rangle .
\label{qft}
\end{equation}

The reversal transformation for QFT can be fulfilled by the following gate 
array.

\begin{picture}(500,310)(0,-30)
\multiput(80,50)(0,50){5}{\line (30,0){30}}
\multiput(140,50)(0,50){5}{\line (30,0){30}}
\multiput(200,50)(0,50){5}{\line (30,0){30}}
\multiput(260,50)(0,50){5}{\line (30,0){30}}
\multiput(320,50)(0,50){5}{\line (30,0){30}}
\multiput(110,40)(60,0){4}{\framebox (30,220)}
\multiput(95,50)(60,50){5}{\circle{10}}
\put(30,200){j}
\put(30,100){k}
\put(60,50){$a_4$}
\put(60,100){$a_3$}
\put(60,150){$a_2$}
\put(60,200){$a_1$}
\put(60,250){$a_0$}
\put(360,50){$b_0$}
\put(360,100){$b_1$}
\put(360,150){$b_2$}
\put(360,200){$b_3$}
\put(360,250){$b_4$}
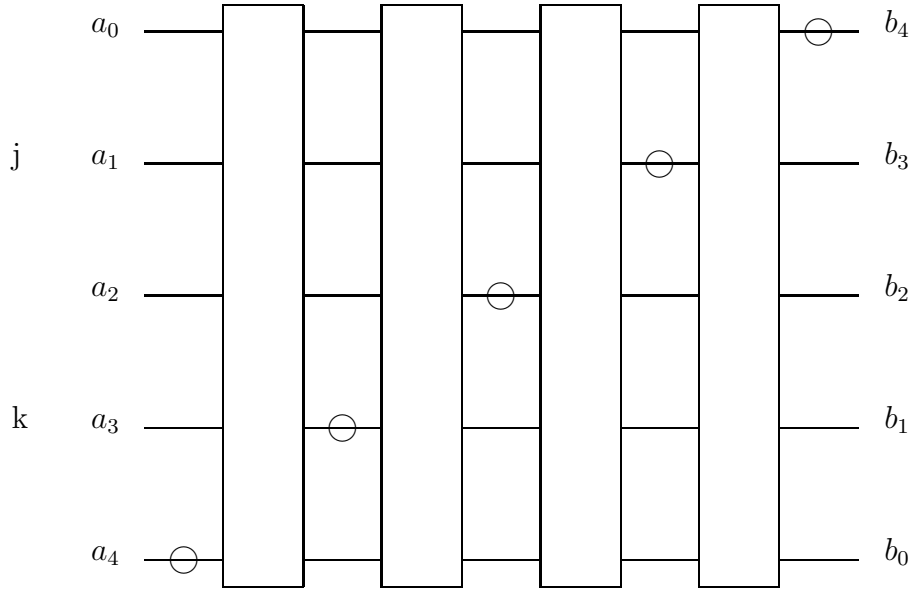
\put(60,10){Picture 2. \  Frameboxes denote fix interaction of the form 
(\ref{Ham}, A),}
\put(120,-5){circles denote Hadamard gates.}
\end{picture}

Here by a framebox we denote a unitary transformation of the form $U=e^{-i\tilde 
H}$ where $\tilde H=\sum\limits_{l>j>k\geq 0}\tilde H_{j,k}$, and each of 
$\tilde H_{j,k}$ has the form (\ref{Ham} , A) with $\rho_0 =\pi$, $r=j-k$. If we 
choose a unit of time so that Plank constant multiplied by $\rho_0$ equals $\pi$ 
and a unit of length so that $r=j-k$ then $U$ will be exactly the transformation 
of state vector induced by the considered Hamiltonian in the unit time frame. We 
assume here that the time of all one-qubit gates action is negligible so that 
two qubits interaction cannot corrupt phases while these gates act. This gate 
array may be obtained from the previous by insertion of "missing" gates 
corresponding to interactions existing physically in the system with constant 
Hamiltonian. To prove that this gate array fulfils $\QFT^{-1}$ we follow the 
method of amplitudes counting proposed in the paper \cite{Sh}. Given a basic 
input state $|a\rangle$ consider the corresponding output state. It is a linear 
combination of basic states $|b\rangle$ with some amplitudes. All modules of 
these amplitudes equal $1/\sqrt{L}$ and we should only count the phases. For the 
simplicity we introduce the notation $a'_j =a_{l-1-j},\ j=0,1,\ldots , l-1$. In 
the process of the gate array application values of qubits with numbers $j$ and 
$k\leq j$ pass through the gates on the picture 2 from left to right. Following 
this passage we separate the following four types of segments: interacting of 
$a'_j$ with itself and $a'_k$ with itself by Hadamard gates, interacting of 
$a'_j$ with $a'_k$ ($j>k$), interacting of $a'_j$ with $b_k$ for $j>k$, and 
interacting of $b_j$ with $b_k$ ($j>k$). The times of these actions are: zero, 
$k$, $j-k$ and $l-1-j$ correspondingly. Summing the deposits of all these 
actions we obtain the resulting phase 
\begin{equation}
\pi\sum\limits_{l>j>k\geq 0}\frac{a'_j a_k k}{2^{j-k} (j-k)} +
\pi\sum\limits_{l>j>k\geq 0}\frac{a'_j b_k (j-k)}{2^{j-k} (j-k)} +
\pi\sum\limits_{l>j\geq 0} a'_j b_j +
\pi\sum\limits_{l>j>k\geq 0}\frac{b_j b_k (l-j-1)}{2^{j-k} (j-k)}.
\end{equation}
Denote the first and last summands by $A$ and $B$ correspondingly. Their 
deposits correspond to the actions of diagonal Hamiltonians on $|a\rangle$ and 
$|b\rangle$ correspondingly. Leave this deposit so far - till the next section. 
Take up a part of sum formed by the second and third summands. After the 
replacement $j$ by $l-1-j$ this part acquires the form
\begin{equation}
\begin{array}{cc}
\pi\sum\limits_{l-1>k+j\geq 0}\frac{a_j b_k 2^{j+k}}{2^{l-1}}+\pi\sum\limits_{l-
1\geq j\geq 0} a_{l-1-j} b_k &=2\pi\sum\limits_{l>k+j\geq 0}\frac{a_j b_k 
2^{j+k}}{2^l} =2\pi S+ 2\pi\sum\limits_{l>k,j\geq 0}\frac{a_j b_k 2^{j+k}}{2^l}=
\\
&2\pi S+ 2\pi\frac{ab}{2^l}
\end{array}
\end{equation}
for some integer $S$. The first summand here does not change the phase and we 
obtain all what is required for $\QFT ^{-1}$ but deposits of $A$ and $B$. 

\section{Correcting of phase shifts}

To cope with the deposit of diagonal summands $A$ and $B$ to the phase we 
present one trick. At first consider only a summand $A$. It consists of addends 
of the form $A_{j,k}=c_{j,k} a'_j a'_k$, where $c_{j,k}$ depends only on $j$ and 
$k$ but not on $a$. Declare $j$th and $k$th qubits separated. We shall apply one 
qubit gate NOT several times to all qubits but separated ones to suppress all 
two qubits interactions excluding interaction between separated qubits. 

At first consider a pair of not separated qubits with numbers $p,q,\ q>p$. Their 
permanent interaction in time frame $\D t$ gives the addend $d_{p,q}\D t\ a'_p 
a'_q$ to the phase where a real number $d_{p,q}$ depends only on how fast does 
interaction decrease and not on $a'_p , a'_q$. For example for a decreasing of 
Yukawa type we have $d_{p,q}=e^{-|q-p|}/|q-p|$. Now invert one of these qubits, 
no matter which, say $q$th by NOT gate. Its state will be $1-a'_q$. Then the 
second $\D t$ period of permanent interaction gives the addend $d_{p,q}\D t\ 
a'_p (1-a'_q)$ to the phase. At last restore the contents of $q$th qubit by the 
second application of NOT. The resulting phase shift of this four steps 
transformation will be $d_{p,q}\D t\ a'_p$ and it depends on the contents of 
$p$th qubit only. Now we can compensate this phase shift by a simple one-qubit
 transformation. 
If we consider a pair of qubits with numbers $p,q$ where one, say $p$th is 
separated and other is not, then we can compensate their interaction by the same 
way using one-qubits operations: two NOTs for $q$ and one phase shift for $p$th. 

Now we should so modify this method that compensate all influence of not 
separated qubits simultaneously. For each not separated qubit number $p$ 
consider the Poisson random process ${\cal A}_p$ generating time instants 
$0<t_1^p <t_2^p <\ldots <t_{m_p}^p <1$ with some fixed density $\la\gg 1$. Let 
all ${\cal A}_p$ are independent. Now fulfil transformations NOT on each $p$th 
qubit in instants $t^p_m$ sequentially. In instant 1 fulfil NOT on $p$th qubit 
if and only if $m_p$ is odd. Thus after this procedure each qubit restores its 
initial value. Count the phase shift generated by this procedure. Interaction 
between separated qubits will be unchanged. Fix some not separated qubit number 
$p$ and count its deposit to phase. It consists of two summands: the first comes 
from interaction with separated and the second - from interaction with not 
separated qubits. Count them sequentially.

1. In view of big density $\la$ of Poisson process ${\cal A}_p$ about half of 
time our $p$th qubit will be in state $a'_p$ and the rest half - in $1-a'_p$. 
Its interaction with a separated qubit, say $j$th brings the deposit 
$\frac{1}{2} d_{p,j}a'_p a'_j +\frac{1}{2} d_{p,j} (1-a'_p)a'_j$ that is 
$\frac{1}{2} d_{p,j}a'_j$. 

2. Consider a different not separated qubit number $q\neq p$. In view of 
independence of time instants when NOTs are fulfilled on $p$th and $q$th qubits 
and big density $\la$ these qubits will be in each of states ($a'_p ,\ a'_q$), 
($a'_p ,\ 1-a'_q$), ($1-a'_p , \ a'_q$), ($1-a'_p ,\ 1-a'_q$) approximately a 
quarter of time. The resulting deposit will be $\frac{1}{4}d_{p,q} [a'_p a'_q 
+a'_p (1-a'_q )+(1-a'_p )a'_q +(1-a'_p )(1-a'_q )]$ $=\frac{1}{4}d_{p,q}$.

A total phase shift issued from the presence of not separated qubits in our 
procedure now is obtained by summing values from items 1 and 2 for all 
$p\notin\{j,k\}$. It is 
$$\frac{1}{2}[\sum\limits_{p\notin\{j,k\}}d_{p,j}a'_j 
+\sum\limits_{p\notin\{j,k\}}d_{p,k}a'_k 
]+\frac{1}{4}\sum\limits_{p,q\notin\{j,k\}} d_{p,q}.
$$
This shift can be compensated by one-qubit operations because the first two 
summands depend linearly on the qubits values and the second does not depend on 
qubits values at all. Thus we obtain a gate with permanent two qubits 
interaction and one-qubit operations fulfilling phase shift to $d_{j,k} a'_j
 a'_k$. If we take time frame $\D t$ instead of unit time in this procedure we 
obtain the phase shift to $\D t\ d_{j,k} a'_j a'_k$. If we want to obtain the 
shift to
$-\D t\ d_{j,k}a'_j a'_k$ we should at first apply NOT to the $j$th qubit, then  
apply the above procedure, then again apply NOT to $j$th qubit and at last add 
$-\D t\ d_{j,k}a'_k$ by one-qubit operation. So we are able to make any addition 
of the form $c\cdot a'_j a'_k$ to the phase for real $c$ independently of its 
sign. An appropriate combinations of such gates gives us a phase shifts to
\begin{equation}
\sum_{j,k} c_{j,k} a'_j a'_k
\label{Phase}
\end{equation}
 for any $c_{j,k}$. Placing such gates before and after $\QFT^{-1}$ procedure
 from the previous section we compensate summands $A$ and $B$ in the phase and 
thus obtain a gate implementing $\QFT^{-1}$. 

Errors arising in this method issue from the possible imperfection of Poisson 
processes generating instants for the qubits inversions and interactions 
continuing in course of these inversions. It can be minimized by increasing of 
density $\la$ and by decreasing of the time needed for NOT operation in 
comparison with typical time of two qubits operation defined by the interaction 
$d_{j,k}$. 

\section{Possibility to use interactions of the different types}

Up to now we considered Hamiltonian of two qubits interaction given by 
(\ref{Ham}, A) with decreasing of Yukawa type. Now we are going to extend our 
technique at first to Hamiltonians of the form (\ref{Ham}, A) for arbitrary 
degree of decreasing, then to Hamiltonians of the form (\ref{Ham}, B) for 
arbitrary decreasing (even for different degrees of decreasing for different 
pairs of qubits), at last - to the interactions which can be diagonalized by 
one-qubit transformations. Namely, we shall prove that it is possible to 
generate a phase shift of the form (\ref{Phase}) and of the form 
\begin{equation}
\sum\limits_{l>j>k\geq 0} c_{j,k}a'_j b_k 
\end{equation}
for any $c_{j,k}$. 

Consider an interaction of the form (\ref{Ham}, A) for arbitrary degree of 
decreasing. Generating of quadratic phase shifts of the form (\ref{Phase}) 
remains unchanged. The only thing we should do is to generate $\QFT^{-1}$ in 
within quadratic phase shift. To do it we start with a gate array represented at 
the picture 2 but now a time interval $\D T$ between nearest Hadamard gates will 
not equal 1 - it will be determined a bit later. Given a number $j$ let $t_j$ be 
instant of the corresponding Hadamard transformation. We shall obtain the 
required gate array inserting sequential NOT operations on each qubit. Construct 
a list of Poisson random processes ${\cal A}_p$ $p=0,1,\ldots , l-1$ 
corresponding to qubits and generating time instances for NOT operations on 
them. But now these processes will not be completely independent. At first 
require that a number of these NOT operations on $j$th qubit preceding $t_j$ is 
even. It will guarantee that a value of $j$th qubit before Hadamard transform 
will be $a'_j$. Then require that $j$th and $k$th processes generate the same 
time instances in the time frame $(t_{j,k} , t_{j,k} +\D t_{j,k} )$ 
corresponding to each pair of qubits $j$th and $k$th, $j>k$ (qubits are 
numerated from bottom to top). Call these time frames ($j,k$)th interval of 
synchronization. They will have the following properties: 

\begin{itemize}
\item[a)] they will not overlap for different pairs, and
\item[b)] each time frame $(t_{j,k} , t_{j,k} +\D t_{j,k} )$ lies between the 
instances of $k$th and $j$th Hadamard gates,
\item[c)] $\D t_{j,k} \gg 1/\la$ for a density $\la$ of processes ${\cal A}_p$.
\end{itemize}
Appropriately choosing $\la$ we can satisfy c). Consider subdivision of 
intervals between nearest Hadamard gates to equal halves and agree that the 
points of division belong to the left of possible time frames. Agree that 
($j,k$)th interval of synchronization lies in the same half as $t=(t_j +t_k 
)/2$. Then only $r<2l$ such intervals will lie in the same half and to make them 
not overlapping we only should divide each half to $r$ segments which lengths 
will be completely determined when we choose the values of $\D t_{j,k}$, 
associate each interval with some segment and then assume that each of these $r$ 
intervals belongs to the corresponding segment. Thus we ensure conditions a) and 
b). Now to determine a gate array (look at the picture 3) we need only values 
$\D t_{j,k}$. 

\begin{picture}(500,290)(0,-35)
\put(25,199){$a'_j$}
\put(25,98.6){$k$}
\multiput(95,78.7)(35,19.9){8}{\line (10,0){10}}
\multiput(100,78.7)(35,19.9){8}{\circle{5}}
\multiput(208,35)(0,10){17}{\line(0,1){3}}
\multiput(218,35)(0,10){17}{\line(0,1){3}}
\put(192,35){$t_{j,k}$}
\put(223,35){$t_{j,k}+\D t_{j,k}$}
\put(420,199){$j$}
\put(420,98.6){$b_k$}
\put(50,199){\line(165,0){158}}
\put(218,199){\line(175,0){182}}
\put(218,98.6){\line(175,0){182}}
\put(50,98.6){\line(165,0){158}}
\multiput(50,50)(5,2.9){8}{\line(5,3){2}}
\multiput(345,23)(5,2.9){8}{\line(5,3){2}}
\put(214,230){\vector(0,-1){30}}
\put(218,230){Interval of sinchronization}
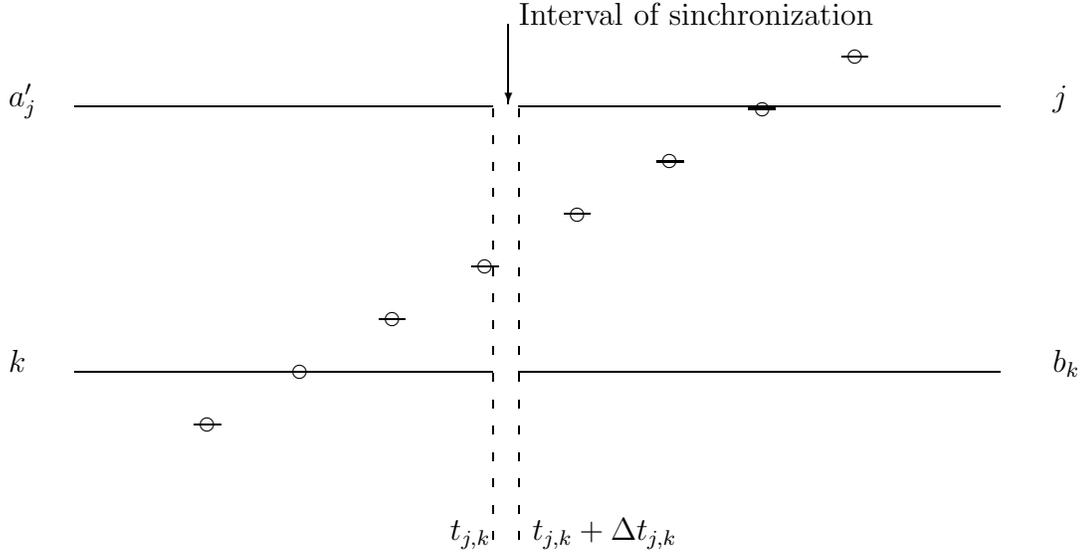
\put(50,10){Picture 3.\ Gate array implementing $\QFT^{-1}$. It consists of one-qubit gates}
\put(100,-5){and different fixed permanent interactions of diagonal form}
\end{picture}

At first consider the case when numbers of NOT operators preceding the first NOT 
in synchronized interval have the same oddity for $j$th and $k$th qubits. Count 
a phase shift generated by this gate array omitting linear summands like $c_j 
a'_j$ or $d_k b_k$ which can be easily compensated by the appropriate one-qubit 
gates: $\D t_{j,k}\frac{1}{2}((1-a'_j )(1-b_k )+a'_j b_k )+S_{lin} +S_{negl}$. 
$S_{lin}$ consists of linear summands, $S_{negl}$ issues from interaction 
outside synchronized intervals and for high density $\la$ it consists of 
mutually canceling additions like above. In the case of different oddities we 
obtain a similar expression. 
In all cases we can choose such values for all $\D t_{j,k}$ (they will depend on 
pairs $j,k$ ) that our gate array fulfils $\QFT ^{-1}$ in within linear phase 
shift and it takes a time $O(l^2 )$. Composing it with gates for appropriate 
linear phase shifts before and after this gate array we obtain a required 
implementation of $\QFT ^{-1}$ in time $O(l^2 )$.

Given interaction of the form (\ref{Ham} ,\ A) where $\rho$ is negative we can 
reduce this case to the considered one by inverting one of interacting gates and 
adding linear shift. It is straightforwardly seen that it results in the only 
change in our constructions as adding linear shift to the phase. By the same 
manner we can implement the direct transformation QFT given a gate array 
implementing its reversal and complexity of the gate array will be about the 
same. 

Given interactions of the form (\ref{Ham},\ B) its application in time frame 1 
gives phase shift to $\rho_1 (1-a'_j )(1-b_k )+\rho_2 (1-a'_j )b_k +\rho_3 a'_j 
(1-b_k )+\rho_4 a'_j b_k $ that may be reduced by the linear phase shift to the 
above case because $\rho_1 +\rho_4 \neq\rho_2 +\rho_3$.

At last this technique may be straightforwardly generalized to interactions with 
Hamiltonians that may be diagonalized by one-qubits transformations. Note that 
this class of interactions is not yet the most general at all, say a model of 
quantum computations with CNOT gates cannot be immediately reduced to it. 

\section{Simulation physics by means of fixed interaction}

Now we take up an important idea dating back to Feynman (\cite{Fe}): to simulate 
physics by quantum computers. A sketch of such simulating method referring to 
Coppersmith-Deutsch-Shor scheme (see picture 1) for $QFT^{-1}$ was proposed by 
Zalka (\cite{Za}) and Wiesner (\cite{Wi}). The method of QFT implementation 
presented above gives an easy way for simulation in case of linear and quadratic 
potentials by means of constant and permanent interaction between qubits. 
Hamiltonians with quadratic potentials serve as a good approximation for 
description of such important physical objects as free particle, ensembles of 
linear harmonic oscillators, free fields, complex molecules. Such Hamiltonian 
for $s_1$ particles has the form
\begin{equation}
H=\sum\limits_{k=1}^s \frac{p_k^2}{2m_k} +\frac{1}{2}\sum\limits_{j,k=1}^s 
v_{j,k}q_j q_k ,
\label{many}
\end{equation}
where $s=3s_1$ is the total number of spatial coordinates $q_k$ determining a 
spatial state of system, $p_k$ are impulses and $v_{j,k}$ are constants. We now 
take up a case $k=1$ because the general case may be considered similarly. 

At first we remind the main ideas of simulation physics by a quantum computer. 
We have to approximate an action of operator $e^{-iHt}$ on a wavefunction 
$\psi_0$ where $H=H_p +H_q$, $H_p =\frac{p^2}{2m}$, $H_q =V(q)$, 
$p=\frac{1}{i}\frac{\partial}{\partial q}$ and potential $V(q)$ is a real 
quadratic function. Without loss of generality we can take $t=1$. To have a 
useful approximation we must deal with coordinate or impulse basis in the space 
of state vectors and in both cases Hamiltonian will not be diagonal. In order to 
reduce the problem to the simple diagonal case choose a small time interval $\D 
t$ and represent our evolutionary operator approximately by
\begin{equation}
e^{-iH}\approx (e^{-iH_q \D t}\ e^{-iH_p \D t})^{1/\D t}.
\label{Ev}
\end{equation}
Choosing, say coordinate basis we have a diagonal operator $H_q$. Using Fourier 
transform $\FT :\ f\ar\frac{1}{\sqrt{2\pi}}\int_{-\infty}^{+\infty}e^{-ipq}f(q)\ 
dq$ and its property to replace derivative $\partial /\partial\ q$ by factor 
$ip$ we can represent an action of impulse part of operator as $e^{-iH_p}=\FT^{-1}\ e^{-ip^2 \D t/2m}\ \FT$ where the medium operator has a diagonal form. If we 
can implement FT and phase shift on $-p^2 /2m$ then the sequential application 
of such operators from (\ref{Ev}) gives the required approximation. 

Assume that a wave function $\psi (q)$ is defined on a segment $(-A,A)$ and its 
impulse representation $\FT\ \psi$ is defined on a segment $(-B,B)$. Choosing a 
small values $\D q$ and $\D p$ we can approximate it by 
$\sum\limits_{a=0}^{2A/\D q} \psi (q_a)\d_a$ where $\d_a (q)$ takes a value 1 on 
a segment $(q_a ,q_a +\D q )$ and zero for other $q$. Then we can approximate FT 
by a linear operator whose action on $\d_a$ gives $\frac{1}{\sqrt{2\pi}}\D 
q\sum\limits_{b=0}^{2B/\D p} e^{-ip_b q_a}\s_b (p)$ where $\s_b (p)$ is one step 
impulse function analogous to $\d_a$. Introducing new one step functions for 
coordinate and impulse by $d_a (q)=\d_a (q-A)$, $s_b (p)=\s_b (p-B)$ we rewrite 
FT in the form 
\begin{equation}
d_a \ \ar\ \frac{1}{\sqrt{2\pi}}\D q\sum\limits_{b=0}^{2B/\D p} e^{-i\ ba\D q\ 
\D p} s_b
\label{FT}
\end{equation}
that looks similar to QFT. 

Assume that the physical space is grained in coordinate and impulse 
representations with the sizes of grains $\D q$ and $\D p$. Then our particle 
may exist only in points of the form $q_a$ or may have impulse only of the form 
$p_b$. We associate a position $q_a$; $\ a=0,1,\ldots , N=2^l$ with the basic 
state $|a\rangle$ of $l$ qubits quantum system. For the simplicity choose such 
units for the length that $\D q=\D p =\sqrt{2\pi}/\sqrt{N}$ and let 
$A=B=\sqrt{\pi N/2}$. Then (\ref{FT}) corresponds to QFT of the form
(\ref{qft}) and phase shift to $-p^2 \D t/2m $ from (\ref{Ev}) corresponds to 
phase shift to $-\pi b^2 \D t /mN $. We can implement the both operations by 
means of fixed interactions because the last has the form from above. At last 
the first operator in (\ref{Ev}) can be implemented by the same way in the case 
under consideration.

If we simulate a system with $s_1$ particles then we should get $s_1$ copies of 
quantum register for one particle and fulfill the procedure described above for 
this joint quantum memory.

\section{Advantages of quantum simulation of wave functions}

A proposed way of quantum solution of Shroedinger equation as well as 
implementation of QFT preserves all advantages of known quantum tricks (look at 
\cite{Sh, Za, Wi}). Namely, given constant interactions between qubits our 
method of implementation of QFT requires the time $O(l^2 )$. If we restrict 
ourselves by approximate version of QFT (it may be obtained by omitting 
exponentially small phase shifts) then the correspondent modification of our 
method takes the time $O(l)$ where a constant depends on the chosen accuracy. 

As for the simulation of wave function the main advantage of quantum method are 
displayed for the case when a simulated system contains many particles. Let we 
have a system of $s$ one-dimensional particles. Its wave function in a fixed
 time instant has the form $\psi (x_1 , x_2 ,\ldots , x_s )$ where $x_j$ denotes 
a coordinate for $j$th particle. To store the approximation of this function 
with grain $\epsilon$ and arguments bounded by $b$ we need of order $N^s$ bits 
where $N=b/\epsilon$. At the same time quantum method of simulation requires of 
order $\log N$ qubits for each particle and all memory will be of order $s\log 
N$ qubits that is logarithm of classical size. But the proposed method shows 
advantages even in case of one particle. Limit the time frame of the required 
simulation for the simplicity by 1, a coordinate and impulse - by $B$. To use 
the formula (\ref{Ev}) we must have only $\D t\ar 0$. Then the total time of 
quantum simulation has an order $\log^2 (1/\epsilon)\frac{1}{\D t}$ whereas an 
approximate solution of Shroedinger equation on a classical computer requires 
scanning of all massive of wave function values that is $1/\epsilon$ in each of 
$1/\D t$ passes. Hence quantum methods of simulation give almost exponential 
time and space saving as compared with classical methods. 

\section{Conclusion}

We considered a model of quantum computer controlled only by one-qubit impulses 
whereas interaction between qubits is fixed and remains unchanged in course of 
computation. Advantage of this model is taken of the simplicity of control. It 
was found a simple way to implement Quantum Fourier Transform by such quantum 
computer. This method makes possible to solve Shroedinger equation for linear 
and quadratic potentials and thus in principle such type of computer can 
simulate systems of harmonic oscillators, free fields and particles and 
molecules consisting of many atoms. 

Further investigations can go to several directions. The first is: to clarify 
possibilities of such simple model of quantum computer, say to find an effective 
implementation of Grover search algorithm (\cite{Gr}) in the framework of this 
model (preliminary step in this direction was made in \cite{FG} but it is not a 
complete feasible solution). Would be important to extend the results to not 
diagonal Hamiltonians. At last it may be interesting to reformulate some areas 
of quantum mechanics in terms of qubit representation of wave function that we 
used for quantum computing. For example such reformulation assumes the existence 
of spatial grain and traditional difficulty of field theory issuing from the 
divergence of rows for high frequencies would be removed. This reformulation 
also can turn to be more economical than the conventional because the time and 
space resources required to describe a many particles system grows much slower 
than for conventional case when the number of particles increases.

\end{document}